# Optimal Sizing of Hybrid Renewable Energy Based Microgrid System

Irfan Rahman
Department of EEE
*Islamic University of Technology*
Gazipur,Dhaka,Bangladesh
irfanrahman@iut-dhaka.edu

Farheen Suha
Department of EEE
*Islamic University of Technology*
Gazipur,Dhaka,Bangladesh
farheensuha@iut-dhaka.edu

Ashik Ahmed
Department of EEE
*Islamic University of Technology*
Gazipur,Dhaka,Bangladesh
ashik123@iut-dhaka.edu

*Abstract—* With the decline of fossil fuel reserves and the escalating global average temperature, the quest for environmentally friendly and renewable energy sources has gained significant momentum. Focus has turned to wind and photovoltaic energy, but their variable inputs necessitate energy storage for reliable power. Economic viability of hybrid renewable power requires meticulous optimization of generating units to ensure uninterrupted and efficient energy production. This paper presents an optimal sizing approach for a Wind-Photovoltaic-Biogas-Battery system using a single objective optimization (SOO) method. A comprehensive comparative analysis is conducted, evaluating the convergence speed and objective mean (for minimization) of seven metaheuristic optimizers: Particle Swarm Optimization (PSO), Aquila Optimizer (AO), Pelican Optimization Algorithm (POA), Dandelion Optimizing Algorithm (DOA), Gazelle Optimization Algorithm (GOA), Zebra Optimization Algorithm (ZOA), and Osprey Optimization Algorithm (OOA). The results demonstrate that the Pelican Optimization Algorithm (POA) outperforms other existing algorithms, exhibiting faster convergence and lower objective mean.

## I. INTRODUCTION

In a world yearning for a harmonious equilibrium, renewable energy sources hold the key to unlocking a brighter tomorrow. They present us with an unparalleled opportunity to meet our ever-growing energy needs sustainably and responsibly[1, 2]. Enter renewable energy sources[3]: the champions of a cleaner, greener future. Solar energy, hydro energy, wind energy, biomass, and biogas have emerged as beacons of possibility, offering efficient and sustainable means of power generation[4]. These sources, when harnessed together in hybrid systems, possess the power to transform our energy landscape, providing us with abundant electricity while treading lightly on the Earth[5]. HRES enable the opportunity of integrating both renewable and conventional energy sources, offering the potential to achieve enhanced efficiency compared to individual power sources[6]. Within the various functioning Hybrid Renewable Energy System (HRES) technologies, those that have proved efficacy, environmental friendliness, and economic feasibility are the combination of Photo-Voltaic (PV) cell, Wind Turbine(WT), Biomass and Battery Storage System (BSS)[7].

## II. BACKGROUND

Sustainable Energy Development Strategies[8] typically focus on three key technological changes: demand-side energy savings, increased energy production efficiency, and the replacement of fossil fuels with renewable energy sources[9]. Large-scale renewable energy implementation plans must include coherent strategies for integrating renewable sources into energy systems influenced by energy savings and efficiency initiatives[10].

Hybrid Renewable Energy Systems(HRES) are a novel power generation method that combines two or more renewable energy sources with traditional ones[11]. Popular optimization methods like Particle Swarm Optimization (PSO)[12], Aquila Optimizer (AO)[13], Pelican Optimization Algorithm (POA)[14], Dandelion Optimization Algorithm (DOA)[15, 16], Gazelle Optimization Algorithm(GOA)[17-19], and Zebra Optimization Algorithm (ZOA)[20, 21] frequently mimic biological behavior.

An optimization case study of an off-grid hybrid system was done in Indonesia with PV panels, a bio generator, a diesel generator, batteries, and the grid optimally addressed energy needs[22]. Research demonstrated the technological advantages of a hybrid PV-BESS for renewable energy utilization and investigated the feasibility of a Building Integrated PV (BIPV) system with and without a battery[23]. For rural electrification a study was done focusing on designing an ideal Hybrid Renewable Energy System (HRES) using solar PV, wind turbines, and bio generators[24]. Load forecasts for residential, commercial, institutional, and agricultural sectors were calculated for reliable electrification[25].

Papers[7, 26, 27] identify PV-WT-BS (photovoltaic, wind turbine, and battery storage) as the most cost-efficient Hybrid Renewable Energy System (HRES) combination, utilizing six optimal sizing methods for configuration determination. In contrast, papers[28-30] explore PV-WT-DG-BS (photovoltaic, wind turbine, diesel generator, and battery storage), employing various algorithms to optimize system sizing.

All these works in the literature review motivated us to study HRES optimized microgrid system. According to the literature review, the most promising combination of HRES components is a PV module and a wind turbine, with the battery serving as energy storage. Thus, in this study, we considered the aforementioned combination, along with biomass, as a new renewable source. Our main objective is to propose an optimization technique for a wind-photovoltaic-biogas-battery hybrid renewable energy system that is both cost-effective and guarantees zero power supply probability. In addition, several recent optimization algorithms for a wind-photovoltaic-biogas-battery hybrid renewable energy system will be compared to get the optimal sizing of the microgrid system.

## III. METHODOLOGY

This section construes elliptically the procedures which were used to achieve the optimum outcome. In this study, the optimum result is characterized as the lowest achievable expenditure for the setup, maintenance, and operational cost of the chosen components of the HRES for a period of 25 years while keeping LPSP close to zero throughout the entire duration. All the calculations done in this study are based on hourly basis.

### A. System Architecture

Hybrid Renewable Energy System(HRES) was established for the replacement of non-renewable energy

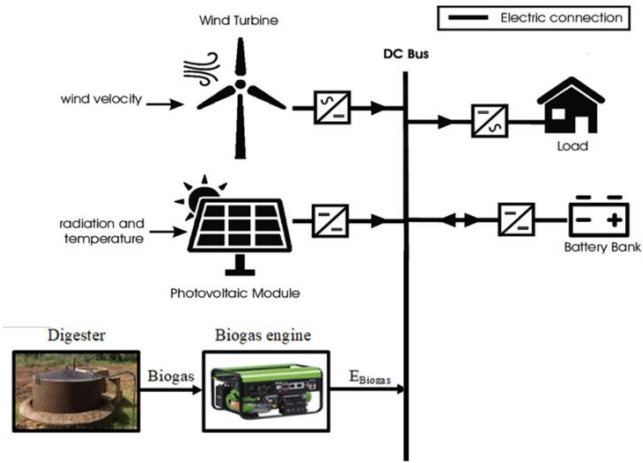

Fig. 1. PV-WT-BG-BS hybrid renewable energy system

sources while fostering the use of renewables. Multiple energy sources are essential for stability due to the unpredictable nature of renewable energy. Such a diversified system offers increased reliability, cost-effectiveness, and efficiency compared to a single-source system[30, 31].

The HRES analyzed in this study consists of four primary components. Three of which are renewable energy sources-specifically, photovoltaic, wind and biomass energy. The fourth component is an energy storage system in the form of a battery, resulting in a PV-WT-BG-BS. These system components are connected to a 24V DC Bus via power electronics converters as shown in fig 1.

For the research, the annual load data for the Islamic University of Technology (IUT), Gazipur has been utilized, employing a Gaussian distribution to derive the hourly load demand.

### B. Wind Turbine Model

Wind turbine, utilized to harness the kinetic energy of the Wind to generate electrical power. The specific power output, $P_w(W/m^2)$, depends on the wind velocity $v(t)$ at that location and is expressed as,

$$\begin{aligned} P_w(t) &= 0 & v(t) < v_{ci} \\ P_w(t) &= av^3(t) - bP & v_{ci} \leq v(t) < v_r \\ P_w(t) &= P_r & v_r \leq v(t) < v_{co} \\ P_w(t) &= 0 & v(t) \geq v_{co} \end{aligned} \quad (1)$$

Where, $P_r$ is the rated power of the turbine, $a = \frac{P_r}{v_r^3 - v_{ci}^3}$ and $b = \frac{v_{ct}^3}{v_r^3 - v_{ci}^3}$. The rated speed, cut-in speed and cut-out speed of the WT are symbolized by $v_r, v_{ci}$ and $v_{co}$ respectively. The following equation is used to calculate wind velocity($v_h$) at a given height[32, 33].

$$v_h = v_r \left(\frac{h}{h_r}\right)^\alpha \quad (2)$$

Here, $h_r$ represents the reference (about 33m)[34, 35]. For this study power law coefficient,$\alpha$ is taken $\alpha = 0.15$, as[36] implies, the site being studied is a decent approximation for such an area because it almost resembles an open topography of grasses. The actual electric power production as obtained from a wind turbine[37, 38] is represented by

$$P_{WG} = P_w A_{WG} \eta_{WG} \quad (3)$$

Here, $A_{WG}$ total swept area by wind turbine and $\eta_{WG}$ refers to the efficiency of the wind turbine generator. The technical specification of the considered WT is presented in table 1[33].

**Table 1.** Specifications of Wind Turbine

| Power (W) | $h_{low}$ (m) | $h_{high}$ (m) | WG capital cost ($) | Tower capital cost ($/unit length) |
|---|---|---|---|---|
| 1000 | 11 | 40 | 2400 | 55 |

### C. Photovoltaic (PV) module model

The power generation of PV modules is influenced by factors besides solar radiation, such as ambient temperature and irradiation conditions, and these characteristics vary from module to module. The output power of a PV module at any one time is determined by the following equation[22, 39]

$$\begin{aligned} P_{PV}(t,\beta) &= N_s . N_p . V_{oc}(t,\beta). FF(t) \\ V_{oc}(t,\beta) &= \{V_{OC-STC} - K_V T_C(t)\} \\ I_{SC}(t,\beta) &= \{I_{SC-STC} + K_I [T_C(t) - 25°C]\} \frac{G(t,\beta)}{1000} \\ T_C(t) &= T_A + (NOCT - 20°C) \frac{G(t,\beta)}{1000} \end{aligned} \quad (4)$$

The expression $P_{PV}(t,\beta)$ denotes the output of the photovoltaic (PV) array during the $t^{th}$ hour, considering a tilt angle of β. $V_{OC}$ and $I_{SC}$ represent the open circuit voltage and short circuit current of a PV module, respectively. FF is the fill factor, while $K_V$ and $K_I$ are the temperature coefficients for open circuit voltage and short circuit current, respectively. G signifies the global solar irradiance on the PV module, TA is the ambient temperature, and NCOT is the nominal cell operating temperature. Following way global solar irradiance can be calculated,

$$\frac{G}{D} = \begin{cases} 1.0 - 0.09K_T & 0 < K_T < 0.22 \\ 0.9511 - 0.1604K_T + 4.388K_T^2 \\ \quad -16.63K_T^3 + 12.33K_T^4 & 0.22 < K_T \leq 0.80 \\ 0.065 & K_T > 0.8 \end{cases} \quad (5)$$

Here, $K_T$ is the hourly clearance index which is the ratio between beam(G) and diffuse(D) components for a slanted PV module.

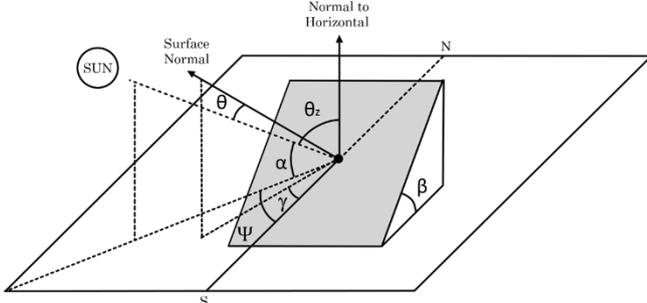

Fig. 2. *Angles related to the sun*

From the figure geometric factor that represents the ratio of beam radiation on a slanted surface to that on a horizontal surface at any given time $R_b$ can be derived. Which is found in the following equation,

$$R_b = \frac{(\cos(\varphi + \beta)\cos\delta\cos\omega + \sin(\varphi + \beta)\sin\delta)}{\cos\varphi\cos\delta\cos\omega + \sin\varphi\sin\delta} \quad (6)$$

Here, φ is the latitude of this site, β is the tilt angle of the PV module. The hour angle, denoted as ω, represents the amount of angular displacement of the Sun. The angle of declination, denoted by δ, represents the Sun's position at solar noon relative to the equator. The declination angle is represented by equation

$$\delta = 23.45 \sin\left(360\frac{284+n}{365}\right) \quad (7)$$

Here, n is the day of the year. Incorporating tilt angle of PV, the total hourly global radiation can be found from the following equation[40] [40]

$$G(t,\beta) = (G-D)R_b + D\left(\frac{1+\cos\beta}{2}\right) + G\rho_g\left(\frac{1-\cos\beta}{2}\right) \quad (8)$$

Here, $\rho_g$ is called the ground reflectance. Total power output from the PV can be determined from the following expression

$$P_{array(t,\beta)} = \eta_{PV} N_S N_P P_{PV}(t,\beta) \quad (9)$$

Where, $\eta_{PV}$ is the efficiency of the converter of pv. The details of the considered PV module are given in table 2[37]

Table 2. Specification of PV module

| $V_{OC}$ (V) | $I_{SC}$ (A) | $V_{max}$ (V) | $I_{max}$ (A) | $P_{max}$ (W) | Capital Cost ($) |
|---|---|---|---|---|---|
| 64.8 | 6.24 | 54.7 | 5.86 | 320 | 640 |

### D. Biogas Modelling

Anaerobic digestion is a valuable waste management process that utilizes microorganisms to break down biodegradable material, producing biogas which can be utilized as a sustainable energy source. The output power of a biogas model is determined using the following equations[41]

$$Gas_{produced} = Food\ waste\ (kg) * Gas_{production\ rate}\ (m^3/kg) \quad (10)$$

$$P_{Biogas} = \frac{V_{Bio} * Cal_{Bio} * \eta_{Bio}}{860} \quad (11)$$

Here, $V_{Bio}$ refers to the Volume of biogas supplied to biogas engine. Calorific value of biogas, Efficiency of Bio engine, Power produced by the Biogas engine are symbolized by $Cal_{Bio}, \eta_{Bio}$ and $P_{Biogas}$ respectively. The specification for the Biogas engine is given in table 3

Table 3. Specification of the Biogas Engine

| Power (W) | Biogas Engine Capital Cost ($) | Digestor volume (m³) | Digestor Capital Cost ($) |
|---|---|---|---|
| 3000 | 720 | 22.183 | 2550 |

### E. Battery Model

The Renewable energy sources are intermittent, creating uncertainty in power generation. To ensure consistent and reliable power supply system, it is essential to have an energy storage system that can store excess energy and release it as required. The charging or discharging state of a battery can be decided through instantaneous sate of charge (SOC). An equation can be used to determine the SOC at any given time

$$I_{bat}(t) = \frac{P_{PV}(t) + P_{WG}(t) - P_{Load}(t)}{V_{bat}(t)} \quad (12)$$

$$SOC(t) = SOC(t-1).\left(1 - \frac{\sigma.\Delta t}{24}\right) + \frac{I_{bat}(t).\Delta t.\eta_{bat}}{C_{bat}} \quad (13)$$

Where, $\sigma$ is denoted as the self-discharging rate of the battery. $\sigma$ is dependent on the cumulative charge and in this study the value is assumed of 0.2%[42]. Furthermore, charging efficiency is fixed at 0.8, while discharging efficiency is set at 1. $I_{bat}$, $C_{bat}$ and $\eta_{bat}$ represents battery current, nominal capacity of the battery and charging efficiency, respectively. $I_{bat}(t)$ created due to the incorporation of the battery where $P_{Load}(t)$ refers to the load demand at the $t^{th}$ hour.

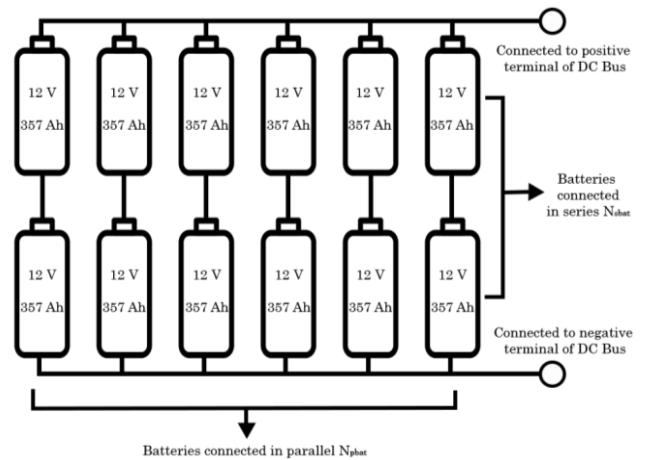

Fig. 3. Batteries connected in series and parallel

The state of charge (SOC) of the battery is critical to maintaining optimal energy balance within the system.

$$N_{bat} = N_{Sbat} \times N_{Pbat} \qquad N_{Sbat} = \frac{V_{BUS}}{V_{bat}}$$

$$C_n = N_{Pbat} \times C_{bat} \qquad (14)$$

The total charge of a battery storage system($C_n$) is determined by the nominal charge of the battery and the number of batteries connected in parallel($N_{Pbat}$). The number of batteries in series($N_{Sbat}$) is dependent on the DC bus voltage($V_{BUS}$), which can be calculated from the equations. To ensure the longevity of battery[43], the concept of maximum charge and the maximum and minimum charging-discharging capacity can be implemented which can be calculated using following equations[44]

$$E_{max} = \frac{C_n \times V_{bat}}{1000}$$

$$E_{cap\_max}(t) = (SOC_{max} - SOC(t)) \times E_{max} \qquad (15)$$

$$E_{cap\_min}(t) = (SOC(t) - SOC_{min}) \times E_{max}$$

Table 4. Specifications of Battery

| Price($) | Voltage(V) | Capacity(Ah) |
|---|---|---|
| 1239 | 12 | 357 |

### F. Objective function Formulation

The objective function is formulated on the basis of loss of power supply probability(LPSP). The economic viability of a Hybrid model is determined by its capacity to fulfill load demand, as indicated by LPSP[44].

The main objective of this research is to reduce the costs associated with HRES. The LPSP is kept at close to zero, ensuring maximum reliability. After that, the cost is computed, resulting in a single-objective optimization.

This study assumes a 25-year lifespan for the HRES under examination. The associated costs include not only the initial setup cost of the PV, WG, and batteries but also the maintenance cost throughout its operational lifespan. The objective function can be expressed as follows

$$\begin{aligned}
Minimize\, f(N_{PV}, N_{WG}, N_{bat}, \beta, h, N_{bio}) \\
= [N_{PV}(C_{PV} + 25M_{PV}) \\
+ N_{WG}(C_{WG} + 25M_{WG} + hC_h + 25hM_h) \\
+ N_{bat}(C_{bat} + Y_{bat}C_{bat}) \\
+ (25 - Y_{bat} - 1)M_{bat} \\
+ N_{bio}(C_{biogenerator} + 25M_{biogenerator}) \\
+ C_{digester} + 25M_{digester})] \qquad (16)
\end{aligned}$$

Subject to the constraints

$$N_{WG} > 0\,; N_{PV} > 0; N_{bat} > 0\,; N_{bio} > 0 \qquad (17)$$
$$90° \geq \beta \geq 0;\, 11 \geq h \geq 40$$

In the objective function, $N_{PV}, N_{WG}, N_{bat}\, and\, N_{bio}$ are the number of PV modules, WGs, batteries and bio-engines respectively, $C_{PV}, C_{WG}, C_{bat}, C_{biogenerator}\, and\, C_{digester}$ are the capital cost of PV modules, WGs, batteries and bio-engines respectively, $M_{PV}, M_{WG}, M_{bat}, M_{biogenerator}$ and $M_{digester}$ are the annual maintenance cost of PV modules, WGs, batteries, bio-engines and digester respectively, $C_h$ is the capital cost per unit height of WG tower, $M_h$ is the yearly maintenance cost per unit height of a WG tower and $Y_{bat}$ is the expected number of battery replacements during the life of HRES.

### G. Hybrid energy system strategy

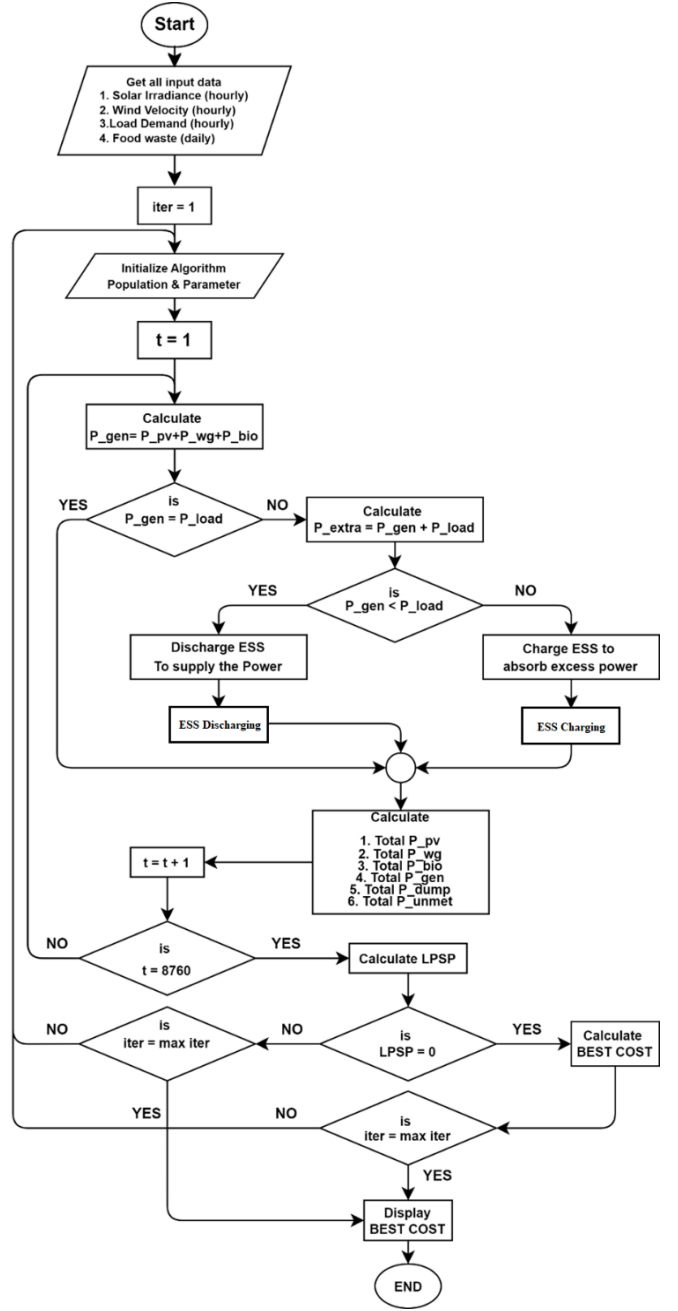

Fig. 4. Flowchart of the operation of a hybrid micro grid system

The optimization process begins by initializing factors such as solar irradiance, wind velocity, load demand, and food waste. The population and algorithm parameters are then set, and six solution sets with defined constraints for variables like $N_{PV}, N_{WG}, N_{bat}, \beta, h\, and\, N_{bio}$ are randomly generated. These sets are fed into the appropriate models to quantify electricity generation. The program determines if overall generation matches demand; surplus energy is stored, while any deficiency is supplied by a battery. If there is an ongoing deficiency, unmet energy is acknowledged. This monitoring continues for a year and culminates in the calculation of the Loss of Power Supply Probability. If the LPSP is 0, the fitness function is computed. The iterative process repeats until the maximum number of iterations is reached.

## IV. RESULTS AND ANALYSIS

This section provides a comprehensive analysis of the superiority of the Pelican optimization algorithm by conducting a comparative assessment against six alternative algorithms. Each algorithm was simulated 30 times in a row, independently. Each of these 30 distinct runs consisted of 300 iterations with a population of 150 to ensure a complete execution of the algorithm.

### A. Analysis of the Algorithms

The primary objective of this study was to identify the most effective optimization algorithm that can achieve the lowest best cost while utilizing limited resources and minimizing expenses. A lower best cost indicates a better optimization algorithm. Analyzing table 5 and Fig 5 it can be said that Pelican Optimization Algorithm (POA) exhibited the lowest best cost of 4276504.73$.

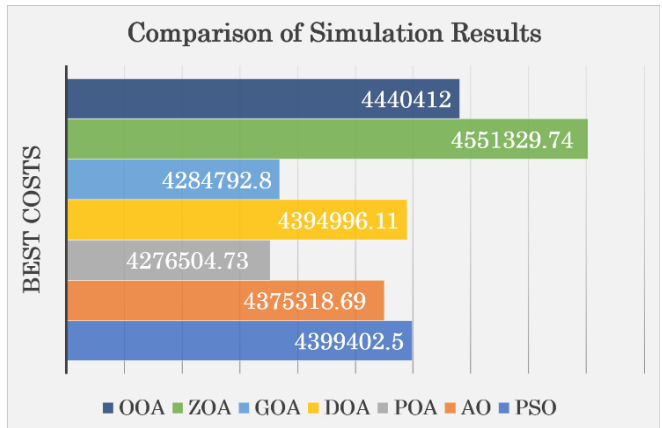

Fig. 5. Comparison of Algorithm by simulated result

This indicates that POA outperformed the other algorithms in terms of achieving the desired optimization goal while utilizing limited resources and minimizing expenses. The POA algorithm demonstrated superior efficiency in finding the optimal solution, resulting in lower costs compared to other algorithms.

Table 5. Performance Comparison of Algorithms

| Optimization Algorithm | Best Cost($) | Difference |
|---|---|---|
| Particle Swarm Optimization | 4399402.5 | 2.87% |
| Aquila Optimizer | 4375318.69 | 2.31% |
| Pelican Optimization Algorithm | 4276504.73 | 0% |
| Dandelion Optimizing Algorithm | 4394996.11 | 2.76% |
| Gazelle Optimization Algorithm | 4284792.80 | 0.19% |
| Zebra Optimization Algorithm | 4551329.74 | 6.43% |
| Osprey Optimization Algorithm | 4440412 | 3.85% |

The Pelican Optimization Algorithm (POA) and Gazelle Optimization Algorithm (GOA) show the lowest percentage differences at 0% and 0.19%, performing nearly as well as the algorithm with the lowest cost (POA). Aquila Optimizer (AO) and Dandelion Optimizing Algorithm (DOA) have slightly higher differences (2.31% and 2.76%), displaying competitive performance. Particle Swarm Optimization (PSO) and Osprey Optimization Algorithm (OOA) have differences of 2.87% and 3.85%, offering considerable optimization capabilities. The Zebra Optimization Algorithm (ZOA) has the highest difference at 6.43%, indicating lower efficiency in minimizing the best cost.

### B. Feasibility Comparison of Algorithms

To evaluate the feasibility of the optimized model by analyzing the time period required for the facility to achieve profitability. Time required for an algorithm to achieve profitability is listed in table 6,

Table 6. Feasibility Comparison of Algorithms

| Optimization Algorithm | Period Required to Be Profitable |
|---|---|
| Particle Swarm Optimization | 21years 2months 12days |
| Aquila Optimizer | 21years 1months 9days |
| Pelican Optimization Algorithm | 20years 7months 16days |
| Dandelion Optimizing Algorithm | 21years 2months 13days |
| Gazelle Optimization Algorithm | 20years 8months 1day |
| Zebra Optimization Algorithm | 21years 11months 16days |
| Osprey Optimization Algorithm | 21years 5months 1day |

From, the table 6 it is prominent that POA becomes profitable after 20 years 7months which about 1 month earlier than GOA. Based on the comparison of the best costs & feasibility the optimization algorithms can be ranked as follows: POA exhibited the lowest best costs & faster time achieve profitability, followed by GOA, AO, DOA, PSO, OOA, and ZOA.

### C. Optimum Sizing of the HRES

Finally, Table 7 gives the optimum HRES combination for the different algorithms. In this table, the configuration shown is only for the best cost obtained in 30 independent runs.

Table 7. Optimal Sizing of HRES

| Algorithm | Best Cost $ | $N_{wg}$ | $N_{pv}$ | $N_{bat}$ | β | h | $N_{bio_{eng}}$ |
|---|---|---|---|---|---|---|---|
| PSO | 4399402.5 | 15 | 125 | 3427 | 18 | 22 | 6 |
| AO | 4375318.69 | 14 | 168 | 3381 | 13.1 | 16 | 10 |
| POA | 4276504.73 | 19 | 69 | 3343 | 31.4 | 11.3 | 9 |
| DOA | 4394996.11 | 18 | 267 | 3338 | 8.2 | 12.4 | 8 |
| GOA | 4284792.80 | 19 | 46 | 3362 | 51.6 | 12.6 | 10 |
| ZOA | 4551329.74 | 12 | 159 | 3547 | 14.4 | 13 | 3 |
| OOA | 4440412 | 63 | 79 | 3367 | 23.6 | 14.7 | 5 |

The number of batteries obtained from table 7 are the batteries connected in parallel, the number of batteries in series will always be two owing to the articular setup that was considered in this study.

## V. Conclusions and Prospects for Future Research

Modern optimization techniques are by their very nature random, which leads to differing sensitivities in different fields. The investigation in this work concentrated on their use in relation to Hybrid Renewable Energy Systems (HRES). It is crucial to recognize that the choice of HRES components cannot be made at random and is dependent on site-specific factors. Several metaheuristic algorithms were applied to find the optimal configuration of the PV, WT, Biomass and battery in the microgrid system. Where POA outperformed all other algorithms in terms of achieving not only the desired optimization goal but also in expeditiously achieving profitability compared to alternative algorithms while utilizing limited resources and minimizing expenses. Furthermore, while the current work employed a single objective optimization (SOO) approach, future studies could explore the benefits of utilizing multi-objective optimization (MOO) techniques. MOO has the potential to provide superior performance by simultaneously considering multiple objectives and generating a range of Pareto-optimal solutions.